\documentclass[sigchi]{acmart}

\usepackage{booktabs} 

\usepackage{graphicx}
\usepackage{listings}
\usepackage{color, colortbl}
\usepackage{dirtytalk}
\usepackage{enumitem}

\definecolor{dkgreen}{rgb}{0,0.6,0}
\definecolor{gray}{rgb}{0.5,0.5,0.5}
\definecolor{mauve}{rgb}{0.58,0,0.82}

\lstdefinelanguage{JavaScript}{
keywords={typeof, new, true, false, catch, function, return, null, catch, switch, var, if, in, while, do, else, case, break},
keywordstyle=\color{blue}\bfseries,
ndkeywords={class, export, boolean, throw, implements, import, this},
ndkeywordstyle=\color{darkgray}\bfseries,
identifierstyle=\color{black},
sensitive=false,
comment=[l]{//},
morecomment=[s]{/*}{*/},
commentstyle=\color{purple}\ttfamily,
stringstyle=\color{red}\ttfamily,
morestring=[b]',
morestring=[b]"
}
\setcopyright{none}

\acmDOI{10.475/123_4}

\acmISBN{123-4567-24-567/08/06}

\acmConference[Conf]{ACM Conference}{Month 2018}{City, Country}
\acmYear{2019}
\copyrightyear{2019}

\acmArticle{4}
\acmPrice{15.00}

\begin{document}
\title{Designing for Democratization: Introducing Novices to Artificial Intelligence Via Maker Kits}
\subtitle{Paper Preprint}

\author{Victor Dibia}
\affiliation{%
  \institution{IBM Research}
  \streetaddress{1101 Kitchawan Road}
  \city{Yorktown Heights}
  \state{New York}
  \postcode{10598}}
\email{dibiavc@us.ibm.com}

\author{Aaron Cox}
\affiliation{%
  \institution{IBM Research}
  \streetaddress{1101 Kitchawan Road}
  \city{Yorktown Heights}
  \state{New York}
  \postcode{10598}}
\email{arcox@us.ibm.com}

\author{Justin Weisz}
\affiliation{%
  \institution{IBM Research}
  \streetaddress{1101 Kitchawan Road}
  \city{Yorktown Heights}
  \state{New York}
  \postcode{10598}}
\email{jweisz@us.ibm.com}

\begin{abstract}
Existing research highlights the myriad of benefits realized when technology is sufficiently democratized and made accessible to non-technical or novice users. However, democratizing complex technologies such as artificial intelligence (AI) remains hard. In this work, we draw on theoretical underpinnings from the democratization of innovation, in exploring the design of maker kits that help introduce novice users to complex technologies. We report on the design of TJBot, an open-source robot that can be programmed using cloud-based AI services. We highlight principles adopted in the design process -- approachable design, simplicity, extensibility, and accessibility -- as well as insights we learned from showing the kit at workshops (N=66 participants) and how users interacted with the project on GitHub over a 12-month period (Nov 2016 - Nov 2017). We find that the project succeeds in attracting novice users (40\% of users who forked the project are new to GitHub) and a variety of demographics are interested in prototyping use cases such as home automation, task delegation, teaching, and learning.
\end{abstract}

%
%


\graphicspath{{figures/}}

\maketitle

\section{Introduction}

\begin{figure}[t]
  \centering
  \includegraphics[width=\columnwidth]{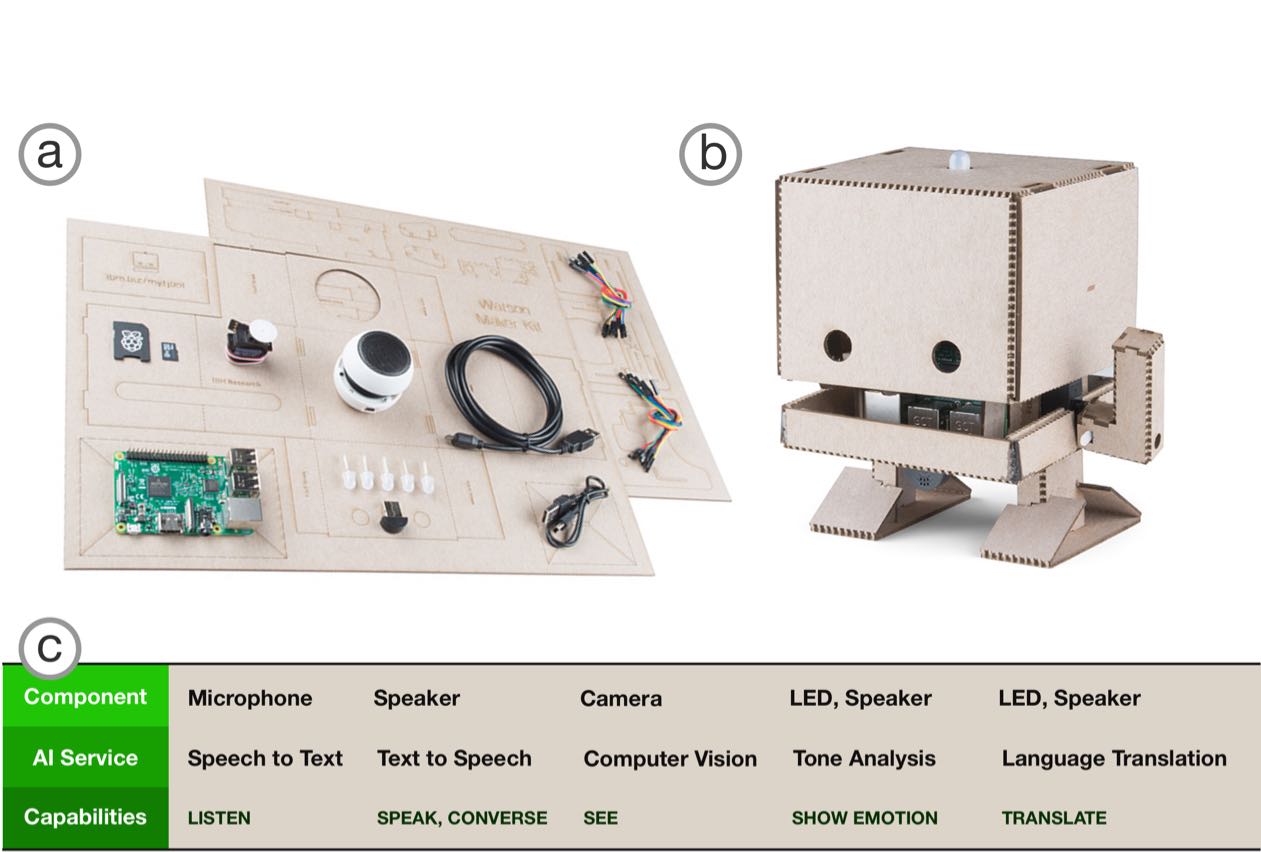}
  \caption{(a) Kit cardboard (chipboard) and components, pre-assembly. (b) Fully assembled kit. (c) Examples of how kit components are combined with AI services to create advanced behaviors.
  \label{fig:tjbotkit}} 
\end{figure}

\begin{figure*}[ht] 
  \includegraphics[width=\textwidth]{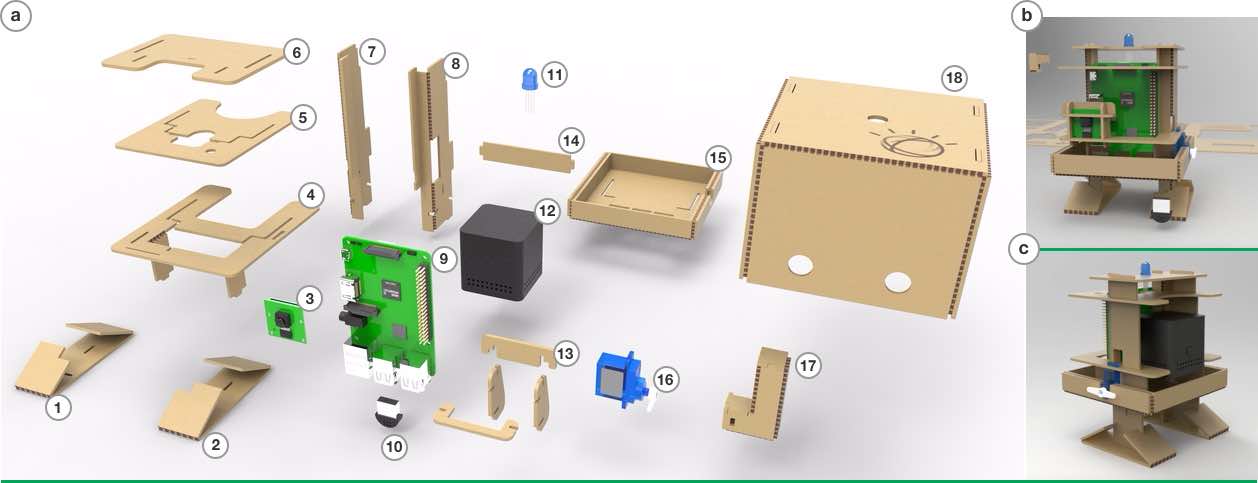}
  \caption{(a) Exploded view of TJBot's components: (1) Left foot, (2) Right foot, (3) Camera, (4) Bottom retainer, (5) Top retainer, (6) LED retainer, (7) Right leg, (8) Left leg, (9) Raspberry Pi, (10) Microphone, (11) LED, (12) Speaker, (13) Camera braces, (14) Leg brace, (15) Jaw, (16) Servo motor, (17) Arm, (18) Head. (b,c) Assembled TJBot with head removed, front and rear view.\label{fig:kitcomponents}}
\end{figure*}

Buoyed by recent advances in machine learning, the general field of AI is well-positioned to solve problems across diverse domains, and it has been referred to as the most important general-purpose technology of our era~\cite{Brynjolfsson2017TheIntelligence}. Speedy declines in the error rates for perception (e.g. speech recognition, image recognition) and cognition tasks (e.g. learning to play complex games such as AlphaGo~\cite{Silver2017MasteringKnowledge}) performed by AI systems herald an era where machines match and outperform their humans counterparts~\cite{He2016,Lake2015Human-levelInduction,Mnih2015Human-levelLearning}. In the most successful forms of its applications, domain experts (e.g. medicine, chemistry, art etc.)  collaborate with AI experts or leverage AI tools in crafting AI-powered solutions. Some examples of such collaborations include AI applied to medical imaging and diagnosis~\cite{Beck2011SystematicSurvival,Rajkomar2018ScalableRecords,Ribli2018DetectingLearning,Geras2017High-ResolutionNetworks}, AI applied to chemical search problems~\cite{Gomez-Bombarelli2016DesignApproach,Ma2015DeepRelationships,Natarajan2016NeuralSurfaces,Sendek2017HolisticMaterials} and AI applied to autonomous vehicle design~\cite{Bertozzi2002ArtificialVehicles}.

While this collaborative model holds promise in further expanding the impact of AI, it is limited by several challenges. The first challenge is related to scale. As a growing discipline, there are not enough AI experts available to collaborate with experts across all domains. More importantly, there is a need to make AI accessible to diverse user groups who can begin to identify problems, assemble data, and create solutions that solely exist within their cultural or social contexts. Second, users without AI expertise may perceive AI, a STEM field, to be complex and unapproachable~\cite{Lyons2012WhenMath, Chilana2015PerceptionsProgrammer,Heilbronner2011SteppingCollege,Nix2015PerceivedFields.}, further deterring them from applying AI to solve their domain problems. These users may consist of students interested in learning, professionals outside the computer science domain (e.g. sales, marketing, medical, chemistry etc.) or individuals familiar with computer science but having no experience with AI (e.g. web and mobile software engineers). Interestingly, while many software developers rank AI as an area of interest, only a few already have the required skill. A recent large scale survey of 101,592 software developers from 183 countries showed that only 7.7\% of developers identified as having skills relevant to AI (data science and machine learning) and ranked AI tools as the third most desired~\cite{Stackoverflow}. Taken together, these challenges necessitate approaches that help on-board and introduce more user groups to AI.


While existing HCI studies have examined the general problem of understanding and supporting a spectrum of novice programmers~\cite{DuBoulay1992ProgrammingNovices,Chilana2015PerceptionsProgrammer,Chilana2016UnderstandingIndustry,Kelleher2005LoweringProgrammers,Myers2009TheHCI}, this work explores the approach of simpler programming language specifications~\cite{Kelleher2005LoweringProgrammers} and the use of visual programming paradigms~\cite{Resnick2009Scratch:All.,Conway2000Alice:Novices} in supporting \textit{learning} goals for users. There is opportunity to systematically enrich the body of HCI theory and design practices through a focus on designing toolkits that make complex technology like AI more \textit{accessible}, as well as studying their impact and limitations in the wild. 

To address this gap, the TJBot project \footnote{TJBot project was created by Maryam Ashoori of IBM Research ~\cite{ashoori2016callingmakers}} -- an open-source maker kit was created
in order to make AI approachable and enable users to easily prototype applications with an embodied agent using pre-built AI services (e.g. speech to text, text to speech, conversation, natural language processing, tone analysis, language translation, etc.). The physical embodiment for TJBot (Figure~\ref{fig:tjbotkit}) can be built from a piece of laser-cut cardboard (Figure~\ref{fig:tjbotkit}a) or 3D printed (see Figure \ref{fig:visualizationsnapshot}c) and contains off-the-shelf electronic components such as a Raspberry Pi, a speaker, servo, microphone, camera and LED. As part of the kit, the team released sample code that demonstrated how to easily combine AI services with the kit's hardware to enable the bot to speak, listen, hold multi-turn conversations, see, translate text, and respond to emotion in spoken words (Figure~\ref{fig:tjbotkit}). These capabilities can then be integrated to build higher-level use cases, such as a storytelling robot for kids, an emotional companion, or a sign language translator using computer vision. Our hypothesis is that by exploiting the learning and engagement benefits \cite{Kuznetsov2011BreakingWorkshops,Rode2015FromMaking,Peppler2013StitchingMaterials,Somanath2017MakerIndia} of maker kits, as well as design principles that make technology approachable, we can create tools that make AI approachable to novice users and support its application in prototyping solutions. 

We acknowledge that democratization is a multifaceted concept with differing implications for different user groups. In this work, our scope of democratization refers to efforts that help make AI more \textit{accessible} to novices and enable its creative \textit{application} in problem solving. As opposed to enabling the creation of complex AI models (e.g. design of novel neural network architectures), the goal is to familiarize users with AI and support its application in a set of problem domains. The target audiences for the maker kit were individuals with some basic knowledge of computing concepts but who have no prior experience with AI. This group included makers~\footnote{Individuals familiar with hardware prototyping and basic programming. AI can serve to enable natural interactions and automation for their projects.}, non-AI software developers, and students. 

To understand the impact of the TJBot kit in the wild, we adopted a multi-method strategy and report on our findings over a 12-month period (Nov 2016 - Nov 2017).
We conducted several workshops (N=66 participants total) and demonstrated the kit at exhibition booths at two large conferences, conducting informal interviews with attendees. Our work contributes to the area of maker kits and their value in democratizing emerging technology in the following ways:  
\begin{enumerate}[label=(\roman*)]
\item we provide one of the first detailed accounts of a design attempt to democratize AI using maker kits, and the strategies pursued in doing so. Our design guidelines  show how to design maker kits that are cheap and easy to use, but highly functional.
\item  we identify a set of use cases of interest for maker kits (home automation, task delegation, teaching and learning), provide insights on specific use behavior (remixing, individual vs group use), and insights on the value of a visualization (interpretable) interface on user interaction.
\item we provide an analysis on user interaction with the sample code for programming the kit (its success in attracting novice users),  challenges faced and limitations to collaborations. 
\end{enumerate}


\section{Background}
\begin{table*}
\centering
\caption{Tools that support the democratization of various AI tasks with estimates of required AI skill.}
\label{tab:aitasks}
  \begin{tabular}{llp{7.5cm}l}
  \toprule
    AI Task     & Description                            & Example Tools                                                &  Skill \\
   \midrule
    Creating AI &              Low level optimization, numeric computation & Python, Numpy                                                & High          \\
    ~           & AI model creation, training, and debugging              &              Tensorflow, Theano, PyTorch, MXNet, Caffe, CNTK & High          \\
    ~           & High Level AI model design                              & Keras, Gluon, Fast.ai, Chainer, Watson Studio, AutoML                                        & Medium          \\
    \midrule
    Applying AI           & Applying prebuilt models (vision, language, etc)                                   & Google, IBM, Microsoft, Amazon, Clarifai                     & Medium          \\
    \rowcolor{lightgray}
    ~           & High-level exploration, embodied prototyping                    & IBM TJBot, Google Vision Kit, Google Voice Kit                                                        & Low         \\
    \bottomrule
    \end{tabular}
\end{table*}

We discuss related work that underpin our research: democratizing innovation via toolkits, maker cultures and maker kits, and AI.

\subsection{Democratizing Innovation via Toolkits}
As new technologies emerge, an important aspect of their long-term success is the degree to which they can be applied to the specific needs of diverse user groups. This vital but challenging component of the innovation process has been described as need-related innovation~\cite[p.147]{vonHippel2006Application:Design}. To address this, research studies have emphasized the user-innovation approach~\cite{Franke2006FindingTheory,VonHippel2005,Morrison2000DeterminantsMarket}, in which companies partner with and co-create with users. 
This approach yields several benefits. First, it enlists a diverse array of external partners, each of whom has a deep understanding of a given usage context and enables the generation of a diverse set of ideas. Next, it enables firms to focus their efforts on identifying highly engaged users~\cite{Hippel2014} and working with them to generate and test concepts~\cite{Urban1988LeadProducts}. User innovation is attractive as it has been shown to enable faster production and reduced costs relative to sole reliance on internal R\&D efforts~\cite[p.148]{vonHippel2006Application:Design,vonHippel2002ShiftingToolkits}. Given the value of user innovation, efforts have been made to understand approaches to enabling user co-creation using toolkits. von Hippel~\cite{vonHippel2002ShiftingToolkits} has discussed the emergence of such innovation toolkits for product design, prototyping, and design-testing tools. These toolkits are intended to enable non-specialist users create high quality solutions that meet their specific needs \cite{vonHippel2006Application:Design}, and thus democratizes the innovation process. To support innovation, von Hippel \cite{vonHippel2002ShiftingToolkits} argues that these toolkits must possess several attributes: support complete cycles of trial and error learning; offer a broad solution space for creativity; offer a friendly user interface; contain reusable modules that can be integrated into designs; and support creation of designs that can be reproduced at scale. Many aspects of the TJBot project followed these design guidelines in order to create an artifact that enabled creativity with a complex technology (AI).




\subsection{Maker Cultures and Maker Kits}
\subsubsection{Making and Engagement} 
In recent years, the culture of making -- also referred to as the maker movement or Do-It-Yourself (DIY) culture -- has moved from being a niche or hobbyist practice to a professional field and emerging industry \cite{Ames2014MakingNot,Lindtner2014EmergingIncubators}. It has been defined broadly as "the growing number of people who are engaged in the creative production of artifacts in their daily lives and who find physical and digital forums to share their processes and products with others"~\cite{Halverson2014TheEducation}. Proponents of maker culture such as Chris Anderson~\cite{Anderson2012Makers:Revolution}, distinguish between the maker movement and work by inventors, tinkerers and entrepreneurs of past eras by highlighting three characteristics: a cultural norm of sharing designs and collaborating online, the use of digital design tools, and the use of common design standards that facilitate sharing and fast iteration~\cite{Anderson2012Makers:Revolution}. Research in this area has sought to understand the formation of online maker communities~\cite{Buechley2008TheEveryone,Buechley2010LilyPadCommunities,Kuznetsov2010RiseCultures} as well as characterize the dominant activities, values, and motivations of participants. Perhaps the most relevant aspect of maker cultures and maker communities to our study is related to the motivations and ethos observed within these communities. Participants have been described as endorsing a set of values such as emphasizing open sharing, learning, and creativity over profit and engendering social capital~\cite{Kuznetsov2010RiseCultures}. Makers have also been described as participating for the purpose of receiving feedback on their projects, obtaining inspiration for future projects and forming social connections with other community members~\cite{Kuznetsov2010RiseCultures}. In general, maker culture promotes certain ethos and cultural tropes such as "making is better than buying"~\cite[p.2604]{Tanenbaum2013DemocratizingPractice} and has been known to build and reinforce a collective identity that motivates participants to make for personal fulfillment and self-actualization~\cite{Somanath2017MakerIndia}. Taken together, these motivations and behaviors suggest maker culture encourages an overall intrinsic motivation approach valuable in addressing known engagement problems \cite{Heilbronner2011SteppingCollege} associated with effortful learning tasks.

\subsubsection{Maker Kits and Learning} 
Maker culture, much like user innovation, has adopted the use of toolkits that support problem solving and fabrication but with a focus on their learning benefits.  Early work by Harel and Papert \cite{Harel1991Constructionism} introduced the theory of constructionism that emphasizes embodied production-based experiences as the core of how people learn. Building on this, learning support tools have been designed that allow for digital construction such as the Logo programming language~\cite{Papert1980MindstormsIdeas} and the Scratch programming language~\cite{Resnick2005SomeKids}, as well as and physical hands-on construction such as the LEGO Mindstorms kits~\cite{Resnick1988LEGODesign}, the LilyPad~\cite{Buechley2008TheEveryone,Buechley2010LilyPadCommunities}, EduWear~\cite{Katterfeldt2009EduWearLife}, the Finch robot kit~\cite{LauwersTom2010DesigningConcepts} and MakerWear~\cite{Kazemitabaar2017MakerWear:Children}. In addition, Google released two maker kits after the release of TJBot in order to provide makers with hands-on AI experiences: the Vision Kit~\cite{google2018vision} and the Voice Kit~\cite{google2018voice}. All of these tools, which we collectively refer to as maker kits, emphasize learning through making and have shown promise in empowering users to create self-expressive and personally meaningful designs~\cite{Kafai2014ATextiles}, improving the perception of computing~\cite{Kafai2014ElectronicSchools} and introducing new user groups to an otherwise inaccessible technology or learning experience~\cite{Buechley2010LilyPadCommunities,Mellis2016EngagingDevices}. The making approach has been cited for its potential to democratize technology, improve workforces, improve technology fluency~\cite{Cross2015ArtsProgram}, improve engagement and participation in education~\cite{Cross2015ArtsProgram,Rode2015FromMaking,Peppler2013StitchingMaterials,Somanath2017MakerIndia,LauwersTom2010DesigningConcepts}, empower consumers and contribute to the economy~\cite{Ames2014MakingNot,Sivek2011WeMagazine}.

\subsection{Democratizing AI}


The research field of AI originates from early efforts to simulate aspects of human intelligences using machines \cite{McCarthy1955AIntelligence}. The domain draws on advances in machine learning algorithms that allow machines to reason, learn, recognize patterns, and understand natural language in manners similar to the human brain.

To scale the impact of AI, research and industry stakeholders have begun to explore approaches that help democratize AI by supporting two types of tasks: \textit{creating AI} and \textit{applying AI}. Table \ref{tab:aitasks} provides a summary of tools that make these tasks more accessible to users, as well as the skill requirements for each task.

Creating AI entails the use of low-level numeric computation functions that are then used to create, train, and debug AI models. These tasks are supported by programming frameworks such as Tensorflow~\cite{Abadi2016TensorFlow:Systems}, Caffe~\cite{Jia2014Caffe:Embedding}, Theano~\cite{Bergstra2010Theano:Python}, and PyTorch, and high level model design frameworks such as Keras~\cite{Chollet2015Keras:TensorFlow}, IBM Watson Studio, and Google AutoML. Typically, the task of creating AI is complex compared to applying AI, and it draws on skills spanning programming, mathematics, statistics, and optimization. To reduce the complexity associated with applying AI in solving problems, AI models are now increasingly offered as black-box cloud hosted services~\cite{Spohrer2015CognitionPerspective}. These services remove the complexity of designing, training, and testing AI models by providing ready to use models that can be accessed over API endpoints. Examples include AI services offered by companies such as IBM, Google, Amazon, and Microsoft.

While being of immense value, these frameworks and services still require considerable skill to use, and may be unapproachable to non-technical users.  In this work, we examine an effort to democratize AI that focuses on supporting individuals interested in \textit{applying} AI (high-level exploration and prototyping).

\section{The TJBot Maker Kit}

\begin{figure*}[ht] 
  \includegraphics[width=\textwidth]{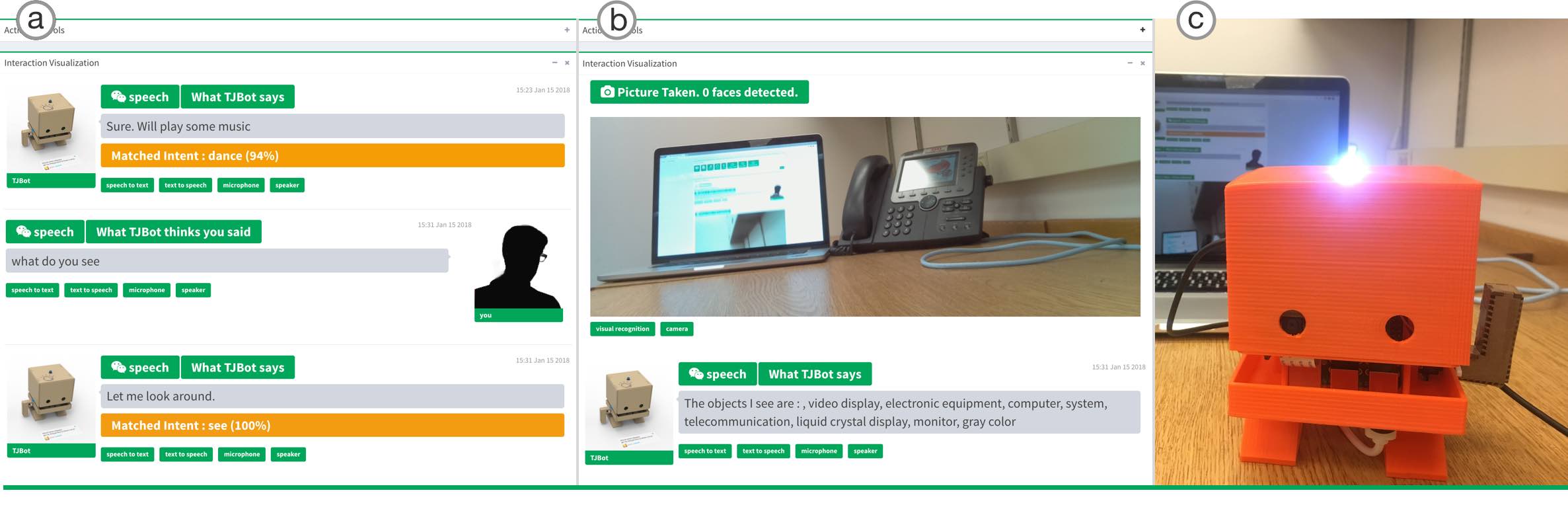}
  \caption{Visualization interface showing: (a) audio transcript, intent confidence values and responses for two conversational interactions, (b) input and response from the computer vision service, and (c) a 3D-printed bot.}
  \label{fig:visualizationsnapshot}
\end{figure*}

We briefly introduce the TJBot maker kit and describe its hardware and software capabilities, early feedback after its release, and emergent design principles identified through discussions with the kit's development team.

\subsection{Design Probe with Developers}
Following the completion of our first prototype \cite{Dibia2017TJBot:Systems}, we created sample code that showed how to program the kit which was then shared on GitHub, an open-source code distribution platform, as well as documentation on how to obtain and assemble the kit hardware. We then showed the kit at a booth within a developer conference where we recruited participants (n=30), and conducted informal semi-structured interviews about the appearance of the kit, its functionality and their overall reaction to demonstrations of the kit. Each of these developers were later sent an early version of the kit, and we monitored GitHub for any issues or feedback they had. While we obtained overall positive feedback during our informal interviews (they felt the kit had an appropriate level of hardware complexity, appeared easy to assemble and felt the sample code was easy to navigate), they had challenges in adapting our sample code to new use cases. We found that users posted issues related to errors when making connections to the prebuilt AI service endpoints, challenges with managing voice interaction context (e.g deciding when to listen or speak during an interaction) etc. While solutions to these challenges are readily obvious to experienced software engineers, many of our users where either novices or were uninterested in solving technical problems unrelated to their primary use cases. To address these challenges, we created a software library api (TJBotlib) which encapsulates capabilities that were complex (usually required a connection to one or more AI services), frequently used and prone to error. This way users could focus on creativity and write significantly less code to realize their use case ideas (see Figure \ref{fig:sampletjbotcode}. We also observed that it was challenging for users being shown a demonstration of the kit to respond to changes in the state of the bot and understand how different cognitive services enabled its capabilities. For example users would frequently ask "what did it hear?", "which of my comments is it responding to?", "why does it give this response? Its not correct". To address this, we began designing a user-friendly dashboard interface (see Figure \ref{fig:visualizationsnapshot}) that would visualize interactions with the bot and make the bot's activities more transparent to the user. Reactions to this interface are reported in the findings section of the paper.

\subsection{Hardware}
The external body of TJBot may be built either from laser-cut chipboard or 3D-printed parts (Figure~\ref{fig:kitcomponents}). The primary electronic component -- TJBot's ``brain'' -- is a Raspberry Pi hardware board, an affordable-yet-functional credit-card sized computer that has become popular within the maker community for prototyping applications. The Pi connects with a set of external sensors and actuators such as an LED, a microphone, a speaker, a servo, and a camera. This hardware enables TJBot to receive visual and auditory input, and generate visual, auditory, and mechanical output.


\subsection{Software}
TJBot is programmed in JavaScript, selected for its ease of use and wide adoption rates~\cite{Meyerovich2013EmpiricalAdoption,Stackoverflow}. It specifically uses the Node.js framework, an open-source cross-platform run-time environment that supports server-side JavaScript development. The Node.js architecture is expressive and functional without sacrificing performance~\cite{Lei2014PerformanceJs,Tilkov2010Node.Programs}, and it has a vibrant, community-maintained repository of 3rd party libraries.

TJBot comes with a set of three stock ``recipes,'' which are pre-written scripts that showcase AI capabilities: changing the color of the LED by voice, performing real-time sentiment analysis on tweets, and conducting a back-and-forth conversation. These recipes enable beginners to interact and experiment with AI services in a no-code/low-code manner; for those with some coding knowledge, the recipes can be tweaked to add new functionality (e.g. making TJBot shine a sequence of colors instead of just one color).

Early feedback on these recipes indicated that they showcased interesting behaviors, but it was difficult for those with some amount of programming knowledge to extend them to perform radically different behaviors. To address this feedback, we developed a software library~\cite{tjbotlib} to hide some of the more difficult aspects of programming with TJBot~\footnote{Other developers have also created their own approaches to simplifying TJBot programming, such as Node RED modules~\cite{tjbot-nodered1, tjbot-nodered2} to enable visual programming.}: authenticating to IBM's Watson APIs, interfacing with the hardware elements, and providing easier asynchronous semantics for features such as listening or speaking.

Another point of feedback centered around \textit{debugging} TJBot's behaviors. Even though the bot has a variety of hardware actuators that may provide feedback, when one's code wasn't working, these actuators did not suffice for debugging. We developed a visual dashboard interface for observing TJBot's internal state~\cite{tjbot-dashboard} (Figure~\ref{fig:visualizationsnapshot}). This interface also enabled us to more easily describe the inner workings of TJBot, and its constituent AI services, to attendees at our workshops and conferences.

\subsection{Emergent Design Principles}
While the design process of TJBot did not begin with explicitly defined design principles, post-design reflection reveals that the following principles were captured in its design.


\subsubsection{Approachable Design}
Designs that are approachable inspire engagement. TJBot inspires people to engage with it via its familiar materials (paper, plastic), its ``tinkerability,'' and its friendly form factor. The square-shaped robot appearance has defined, recognizable features such as two eyes, a mouth, and an arm, conveying its ability to see, speak, and wave.


\subsubsection{Simplicity (Low Floors, High Ceiling, Wide Walls)}
This design goal necessitates a kit that is easy to both physically assemble and program (low floors), yet functional enough to support the creation of complex and meaningful use cases (high ceilings). In this sense, the kit should also support a wide range of user skill levels, from novice users (e.g. school students) to experienced developers. This design goal is also related to the ``wide walls'' approach described by Resnick et al.~\cite{Resnick2005SomeKids} in which they emphasize that construction kits (for kids in their case) should support a wide range of different kinds of exploration (wide walls). The expectation is that a diversity of possibilities will allow for the construction of unique creations that surprise both the user, and the creators of the kit, and inspire that sense of infinite possibilities necessary for sustained engagement.

We believe TJBot achieves this goal in several ways. First, its assembly approach allows users to build the kit by simply folding chipboard or snapping plastic parts together, without the need for complicated gluing. In addition, assembly can be performed by very young children under supervision, resulting in a fun artifact that they can make their own via decoration; no code or AI required. For those with more advanced skills, even the wiring of the electronics can be performed simply, without soldering. Finally, the software developed for TJBot hides much of the complexity of interfacing with hardware and cloud APIs, and enables one to program at the level of the bot's capabilities: speak, see, listen, shine, wave, etc.


\subsubsection{Extensibility}
Extensibility allows users to easily add new capabilities to an existing artifact. As TJBot is an open source project, numerous projects have been developed to give TJBot new abilities, such as walking, driving, speaking different languages, and even communicating with a companion robotic dog.


\subsubsection{Accessibility}
TJBot was designed to be relatively cheap, easily disseminated, and widely available. The use of cardboard as a design material, as well as the use of off-the-shelf electronic components, make it easy for anyone to assemble their own kit. Accessibility is increasingly important for DIY kits as research has shown that in some cases, while maker culture strives to promote democratization of technology, resource constraints (custom components, expensive parts) make it remain an activity of privilege, accessible to only a select few~\cite[p.2605]{Tanenbaum2013DemocratizingPractice}.


\section{Study}
We report on two types of analysis we conducted: informal feedback received from attendees of TJBot workshops and demonstrations, and findings from an analysis of data on the project's Github repository.

\subsection{Workshops and Demonstrations}
We conducted five workshop sessions (N=66 participants total) at two large technology-oriented conferences. We invited participants of all skill levels interested in learning about and building AI-enabled prototypes. Participants were recruited on a first-come, first-served basis; emails were sent out to conference attendees inviting them to sign up until all spaces had been filled. Each session lasted an hour, and participants worked in groups of four. Participants filled out a pre-workshop survey that asked about their background, level of technical expertise, their interest in working with the kit's hardware (sensing) capabilities and the software (AI) capabilities, and the use cases they envisioned. To reduce setup time and ensure each group could write code and interact with the bot, they were provided with a fully-assembled bot preloaded with starter sample code. 

Each session began with a 15-minute introduction to the kit describing its components, the AI services available for use, and the online location of all materials (e.g. design files, sample code, instructions). This introduction also included a demonstration of the kit, showing off AI services such as speech to text, text to speech, computer vision, and sentiment analysis. During the next 20 minutes, each group of participants was guided through a hands-on assembly of the laser-cut chipboard. Finally, participants were instructed in running the sample code on the fully-assembled TJBots, and given an opportunity to examine and experiment with the source code. At the end, participants filled out a post-workshop survey in which they were asked to evaluate the kit's visual appeal, ease of use (programming and assembly), and describe potential projects they might like to create using it. All participants were given a TJBot kit to take home.


\begin{table}[]
\centering
\caption{Themes from reported use cases before and after the workshop.}
\label{tab:usecases}
\begin{tabular*}{\columnwidth}{llll}
 \toprule
\textbf{Theme}             & \multicolumn{3}{l}{\textbf{ \% of use case comments}} \\
 \toprule
                  & Before        & After       & Overall       \\
  \midrule
Home automation   & 9             & 13          & 19            \\
Task delegation   & 28            & 18          & 40            \\
Teaching/Learning & 19            & 10          & 25            \\
General functions & 6             & 37          & 42            \\
None              & 30            & 18          & 41\\
\bottomrule
\end{tabular*}
\end{table}

\subsubsection{Survey Findings}
The pre-workshop survey revealed some diversity among the participants: 51\% identified themselves as developers, 22\% as makers, and 13\% as designers. While most (70\%) had over three years of programming experience, only 26\% had a year or more experience working with embodied maker kits and most (74\%) had no experience with AI services. This population reflects similar trends found in a large scale study of developers where only 7\% of over 100k developers had machine learning or AI skills~\cite{Stackoverflow}. None of the participants had any experience with TJBot, although about a third of them had previously heard about it online. At the end of the workshop, 90\% of participants indicated they were interested in working with AI services going forward.

\textbf{How do users envision their use of the kit?}
Participants reported that they intended to use the TJBot kit in a number of ways. Over half of them (62\%) indicated that they intended to modify the kit's hardware (e.g. adding new types of sensors and actuators), 77\% intended to create additional software components, and about half intended to pursue both kinds of expansions. It is interesting to note that most participants said they would use the kit with other people: 65\% with friends or colleagues, 48\% with children, and 18\% with students (teaching). A minority (27\%) indicated that they would only work alone. In the post-workshop survey, the large majority of participants provided positive feedback about the kit: 96\%, rated it as visually appealing, 96\% felt it had a good repertoire of functions, 86\% felt it was easy to program, and 78\% felt it was easy to build.

\textbf{What are the use cases of interest?}
Participants were asked to describe the use cases they envisioned for TJBot at the start and at the end of the workshop. Using an iterative coding approach, the use cases provided by participants were coded into three main themes and one "General Functions" theme as shown in Table \ref{tab:usecases}. Some participants provided several ideas that fit within multiple themes - these ideas were coded separately.
\\ \textit{\underline{Home Automation:}} One of the most common use cases mentioned was related building a robot that would monitor objects/activities or manage smart devices already present within a home. These use cases focused on using the bot as a voice-based interface that can be used to instantiate monitoring activities as well as report on the status of managing activities.
\\\indent \say{\textit{twitter alert each time a deer walks through my yard}} - P32 \\\indent \say{\textit{home automation assistant (lights, nest), face recognition/welcome sensor}} - P44 \\ \say{\textit{control home automation sensors to help handicap or elderly people}} - P52; \\\indent \say{\textit{Interface to my home automation system, and my personal weather system}} - P37. \\\textit{\underline{Task Delegation (Anthropomorphizing):}}
This theme referred to use cases where participants referred to TJBot as an individual entity, using words like "a friend", "assistant" or "helper". These participants often described their use cases using words that would typically describe interactions with a trusted acquaintance. 
\\\indent \say{\textit{My new best friend}} - P4; \\\indent \say{\textit{I can picture an Alexa-esque friend who can answer questions and do fun things}} - P18.
Some participants described TJBot as a ``personal assistant.'' While this suggests less intimacy than the more personal framings above, it nevertheless indicates a willingness to delegate actions or responsibilities to the bot.
\\\indent \say{\textit{a friend that will greet guests at my door}} - P13; \\\indent \say{\textit{house manager}}- P64; \\\indent \say{\textit{intelligent house robot}} - P54 
\\ \textit{\underline{Teaching and Learning:}} Participants indicated various ways in which they would use the kit for teaching and learning. Several users mentioned they would use the kit either for self-learning, or as a tool to teach their kids (and other young individuals) about artificial intelligence and computer science in general.
\\\indent \say{\textit{just learning machine programming, learn about Watson services}}- P26; \\\indent \say{\textit{using it as a way to teach my kids (10 and 12) about coding with Watson}} - P5; \\\indent \say{\textit{projects with my daughter}} -  P30; \\\indent \say{\textit{getting kids interested in technology}} - P51; \\\indent \say{\textit{would like to use this as a tool to work with young girls learning about technology}} - P60; \\\indent \say{\textit{challenging my kids with development}} P67;
Finally, others saw the kit as an opportunity for professional learning, a tool for communicating technical ideas to non-tech-savvy audiences (clients in some cases) and a tool for developing engaging proof of concepts (POCs).
 \\\indent \say{\textit{I am in the automotive industry. I can definitely use this for proof of concepts.}}- P57
\\ \textit{\underline{General Functions:}}
Participants also came up with many other use cases and ideas that were more related to specific capabilities of the kit - e.g. ideas around the camera, audio, and general input output capabilities of the kit. Examples of these included using the bot as an announcer (announce tennis scores, weather changes, software project build status, security incidents, baby monitor, family greeter), vision recognition tool (recognize foreign currencies, analyze images to infer human activity or identity), a digital assistant for language learning, a tool for games and entertainment, virtual pets and a tool for prototyping interactive stories. 

\subsection{TJBot Demonstrations}
At each conference where we ran the workshops, we also presented TJBot at an interactive booth. Visitors to the booth could interact with an assembled version of the kit, ask questions, and provide feedback. Each demonstration lasted about 10 - 20 minutes and allowed us observe reactions to the kit, and assess desire for specific use cases.


\subsubsection{Reactions to the Kit}
We observed 3 distinct reactions common to most users: an affective reaction in which users reacted to the visual appearance of the bot, a functional reaction in which users inquired about the capabilities of the bot or tried to infer this themselves, and a customization reaction in which they sought to learn about how they could customize the bot to use cases of interest to them.  
\\(i) \textit{Affective}: People tended to anthropomorphize the bot and perceive it as friendly, using words like ``cute'' and ``cute little guy'' to describe it. 
\\(ii) \textit{Functional}: People asked about the bot's hardware and general capabilities and asked about concrete use cases for which it could be used. We also observed interesting assumptions made by users as they tried to interact with the bot. For example, several users would immediately attempt waving at the bot (assuming it could see) and issuing voice commands (assuming it could hear) even before they confirmed the bot was capable of these capabilities. These assumptions were likely inspired by the humanoid appearance of the bot.
\\(iii) \textit{Customization}: People sought to understand the ways that they could extend the capabilities of the bot and understand the complexity of integrating 3rd party hardware and software components.

\subsection{GitHub Activity}
Several collaborative features of GitHub enabled us collect data on usage patterns: issues, stars, and forks. GitHub issues allow people to create bug reports, feature requests, enhancements, or submit general feedback. Project maintainers may also use issues to keep track of open tasks. GitHub stars allow people to bookmark projects of interest and show appreciation to their maintainers. Forks allow people to create a copy of a project that they can modify without affecting the original, and enables them to contribute their modifications if they chose to. During the period of 12 months considered in this analysis, the project repository was forked 173 times and starred 314 times by users. Of those who forked the repository, 40\% had owned a GitHub account for a year or less and had only a single code repository listed on their account. Users also opened a total of 47 issues mainly around requesting support on hardware or software related errors they encountered (68\%), suggesting contributions to the project documentation including identifying broken links or missing content (11\%) and providing general feedback on issues they encountered or fixes they had implemented (21\%). A total of 6 pull requests (GitHub's mechanism for allowing external users contribute to a code repository) were created, of which 4 were approved and merged into the main code repository.   

\section{Discussion}

\subsection{Maker Kits as Tools for Democratizing Technology} 

TJBot addresses its goal of democratizing AI in several ways: the degree to which it elicits interest from users, attracts novice users, and engages broad audiences.  

\subsubsection{Eliciting Interest}
Results from our survey and demonstrations suggest that TJBot is successful in eliciting interest in AI. While 74\% of workshop participants noted no previous experience with AI services, 90\% mentioned they were likely to create AI prototypes after the workshop. We also observed a 3-step reaction when we demonstrated the kit to users: a positive affective reaction to the visual appearance of the kit, exploration of its capabilities, and finally an expression of interest in extending these capabilities to meet personal use cases. In our workshop survey, participants also expressed interest in extending the hardware and software capabilities of the kit.

\subsubsection{Attracting Novice Users}
Attracting individuals to experiment with complex technology like AI can be challenging especially for first time users with limited technical skill. Interestingly, interaction pattern data from GitHub suggests that a significant amount of activity comes from relatively new users. We find that 40\% of users who made public forks of the code had owned their GitHub accounts less than a year and had not authored any other publicly shared projects.

\subsubsection{Impact} 
Although it is difficult to estimate the exact number of total users and programs designed around TJBot (the project does not include any explicit tracking), initial data from social media and other sources suggests the project has been widely used across across a range of demographics and locations. Over the period of 12 months, instructions for the project on Instructables were viewed over 91,000 times, the TJBot library used to program the kit was downloaded over 5,300 times and the project repository on GitHub received over 350 stars. Users also posted images and videos of their prototypes on Twitter spanning use cases such as connecting the bot to their IoT home devices, interactive storytelling and a range of voice based conversation examples, many in agreement with the use cases elicited during our workshop survey. We also saw groups use the kit during tech meetups, corporate and informal training, hackathons, and numerous STEM education events. The variety of interactions on Twitter spanned users from over 10 countries.

\subsubsection{Exploring the Limits of AI}
Recent studies indicate the average user incorrectly estimates the limits of AI~\cite{Chandrasekaran2017ItMind} by overestimating its capabilities or artificially discounting its value. These estimations can foster polarizing (utopian or dystopian) views of AI. A lack of experience with AI has been shown to be partly responsible for this failure to correctly predict what AI can or cannot accomplish~\cite{Chandrasekaran2017ItMind}. Maker kits like TJBot can help address this challenge by providing a medium which enables user to experientially probe the limits of AI. The maker kit enables users to iteratively formulate and test hypothesis on AI capabilities, helping them build an objective assessment on the performance of AI. During these explorations, users can also address limitations they observe. For example, a user may a assemble a new image dataset of faces and train a custom AI model which improves image recognition performance for their skin tone. These activities can both aid learning and serve to increase the diversity of training data and problems that can be addressed using AI.

\subsubsection{Building Trust Through Transparency}
We found that explicitly visualizing the   
input/output of AI services used to prototype an interaction helped users to make sense of the decisions made by TJBot. 
The feedback provided by the dashboard visualizations helped users understand when the system was failing (e.g. incorrect transcripts of a command or keyword, network delays) and helped inform steps to recovery (e.g. retry the interaction or pause).  This observation is similar to results found by Kulesza et al.~\cite{Kulesza2015PrinciplesLearning} where they find that an implementation of an explanatory debugging interface increased user understanding of a machine learning system and allowed a more efficient correction of mistakes. Existing research also suggests that being able to understand how AI systems work is critical for building trust~\cite{Lipton2016TheInterpretability}. Trust is critical as it has been shown to drive system usage. Ribeiro et al.~\cite{Ribeiro2016WhyClassifier} note that users are unlikely to use a system which they do not trust. With the proliferation of integrated smart appliances infused with AI (e.g. connected homes with smart TVs, refrigerators, lighting systems, etc.), our findings suggest that the user experience can be improved by creating visualizations which offer information on the dynamics (input and output state change) of integrated AI services.


\subsection{AI Maker Kit Use Cases and Behaviors}

Results from our survey provide insight on use cases of interest to users and the value of embodied maker kits. 

\subsubsection{Use Cases}
The most common use cases mentioned by participants were task delegation (40\%), teaching and learning (25\%), and home automation (19\%). Based on the study results, we find that participants, perhaps in an indirect reference to the AI capabilities within the kit, envision use cases where the bot can take on the role of a ``manager'' that both performs its own sensing functions and also manages other devices to orchestrate higher level actions within a home environment. This result agrees with extant studies of DIY communities where 35\% of survey respondents indicated they worked on DIY projects for home improvement~\cite{Kuznetsov2010RiseCultures}. They also suggest that users perceive value in AI agents capable to managing tasks on their behalf with important implications for the design such agents. These include robust natural language interaction capabilities, integration capabilities that allows it interface with other agents or systems and reasoning capabilities to intelligently orchestrate high level decisions.

\subsubsection{Social Making}
Respondents indicated that they planned to use the maker with others: with friends or colleagues (65\%), and with children in an educational or family setting (48\%). While further research is required to the drivers of this social/group use behavior, the current findings suggest that such maker kits are suitable for team training, education and parent-child teaching use cases. They further position maker kits as tools to introduce students to AI, with potential to address engagement challenges~\cite{Heilbronner2011SteppingCollege,Watkins2013RetainingMajors} known to deter students from STEM education. A design implication of this finding is the need to explore programming interfaces that support multiple concurrent users and user pairs (e.g. student-student, teacher-student, parent-child, etc). As advances in AI algorithms continue to drive the proliferation of AI, it is expected that more firms will provide black box AI services as well as professional and consumer development kits to support creativity with AI (e.g. \cite{AIY2018Do-it-yourselfIntelligence}). The insights from our experience demonstrating TJBot and surveying users (use cases, social making) can help inform the design of such kits.


\subsection{Challenges with Maker Kits}
Issues reported on GitHub suggest that while most users are able to assemble the kit and download the sample code we provide, they face the most difficulty troubleshooting hardware and software problems during installation and prototyping solutions. These included difficulties correctly issuing some command line instructions and navigating project directories, modifying software configuration to match changes in their hardware components and connectivity issues with cloud hosted AI services. We find that 68\% of all recorded issues on the project's GitHub repository are focused on requesting technical support while only 21\% were focused on providing feedback or solutions. Given that most users were not technical experts, there were only 6 external code contributions to the project on GitHub. This observation highlights the importance of allocating resources to resolving technical issues as an important aspect efforts in democratizing technology. 

To address these issues, a dedicated team member should monitor issues posted on GitHub and continuously provide support, in addition to developing instructional materials such as videos and guides. Automated tests and troubleshooting scripts can help support self-troubleshooting. Troubleshooting hardware issues, such as incorrect wiring or broken components, were much more difficult to perform online via GitHub, and further work is needed to find good solutions to help people accomplish this.

%
%
\subsection{Limitations and Future Work}
We evaluated our kit using surveys and observations within a workshop setting. While this approach is common in maker kit research~\cite{Jacobs2014Dresscode:Making,Katterfeldt2009EduWearLife,Kazemitabaar2017MakerWear:Children,Lau2009LearningStudents,Meissner2017Do-It-YourselfDisabilities}, it limits our ability to assess the effect of a workshop facilitator and workshop content on interaction behaviors. Further experimental research is needed to independently compare the maker kit approach to other approaches (for example, personal fabrication~\cite{Mellis2016EngagingDevices}) of introducing novice users to new technology. Another limitation of this work is related to the sample of interaction data analyzed from GitHub. It is possible that not all new users who interacted with our project (built the kit and downloaded sample code) took the additional effort to create GitHub accounts or public forks limiting our ability to assess this population.      


From our workshops, we have received early feedback from educators eager to integrate the kit into their curriculum. However, they have noted that further simplification is required to make the project accessible younger users (e.g. 6+). This is similar to findings from~\cite{Mellis2016EngagingDevices} where they call for tools that simplify programming components associated with the task of personal fabrication. Thus, future work is needed to focus on further automating aspects of the kit's software and hardware setup, and creating visual programming interfaces (e.g. \cite{Resnick2009Scratch:All.,Cross2013AEducation}) that allow younger or lower-skilled users to prototype solutions. 

\section{Conclusion}
We described how the TJBot maker kit was designed to democratize the emergent area of AI.  We report on principles demonstrated through the kit's design, feedback received from workshops and demonstrations, and insights from analyzing data on how users interacted with its open-source code over a 12-month period. We find that users are interested in exploring a variety of AI use cases including home automation, task delegation, teaching, and learning, that they view working with the kit as a social endeavor, and that the kit has a wide appeal to novices. This work contributes to the area of designing for democratization and proposes maker kits as a viable approach. As more designers and researchers begin to explore the use of cardboard-based construction kits (e.g. Google AIY~\cite{google2018vision,google2018voice}, Nintendo Labo \cite{AIY2018Do-it-yourselfIntelligence,Nintendo2018NintendoSite}), the design principles we present can provide directions on how to design such kits.

\section{Acknowledgement}
We thank Maryam Ashoori whose efforts as the team lead  of the TJBot project was critical to making the project a reality. We thank Rachel Bellamy for her enthusiastic support of this project, and are grateful to Thomas Erickson and Shari Trewin for valuable feedback on this manuscript.

\bibliographystyle{ACM-Reference-Format}
\bibliography{Mendeley.bib,bibliography.bib}

\end{document}